\documentstyle[11pt]{article}
\begin{document}

\centerline{\large\bf Test of CPT symmetry in cascade decays
	\footnote{Supported in part by National Natural Science 
	Foundation of China } } \vspace{1cm}

\centerline{ Dong-Sheng Du, ~~
             Zheng-Tao Wei }

\vspace{1cm}

\begin{center}
CCAST(World Laboratory), ~~P.O.Box $8730$,
Beijing $100080$, China 

and

Institute of High Energy Physics, P.O.Box $918(4)$, Beijing
$100039$, China
\end{center}

\vspace*{0.3cm}

\begin{center} \begin{minipage}{12cm}

\noindent{\bf Abstract}

Cascade mixing provides an elegant place to study the
$B^0-\bar{B^0}$
mixing. We use this idea to study the CPT violation caused by
$B^0-\bar{B^0}$ mixing. An approximation method is adopted to
treat the two complex $B^0-\bar{B^0}$ mixing parameters $\theta$
and $\phi$. A procedure to extract
the parameters $\theta$ and $\phi$ is suggested.
The feasibility of exploring the CPT violation 
and determining of $\theta$ and $\phi$ in the future 
B-factories and LHC-B is discussed.

\end{minipage} 
\end{center}

\vspace{1.5cm}

\newpage

\baselineskip 24pt

\section*{I.  Introduction}

The violations of the three discrete symmetries C, P, T 
have changed our ideas about the physical world. The discovery 
of CP violation 
in $K^0- \bar{K^0}$ complex was made long time ago. Recently, 
The direct T violation \cite{CPLEAR1} 
and direct CP violation ($Re(\epsilon'/\epsilon)\neq 0$) 
\cite{KTeV} are found experimentally. 
Only the combined CPT symmetry is left unbroken. 
The CPT theorem is the  general result of the local, relativistic  
field theory. If it is violated, it will cause the 
fundamental crisis of our present particle theory. The recent tests 
of CPT violating effects give the bounds \cite{CPLEAR2} 
$r_{_K}\equiv
\mid \frac{m_{\bar{K^0}}-m_{K^0}}{m_{K^0}}\mid \leq 10^{-18}$ 
and the very uncertain values 
$Re(\delta)=( 3.0\pm 3.3_{stat}\pm 0.6_{syst})\times 10^{-4}$, 
$Im(\delta)=(-1.5\pm 2.3_{stat}\pm 0.3_{syst})\times 10^{-2}$.  
The tests of CPT symmetry in B system have been suggested from the 
theory \cite{Kostel} and phenomenology \cite{Sanda1} \cite{Xing}. 
In \cite{Sanda1}, the authors point out that CPT violation in 
$B^0\bar B^0$ mixing can lead to a dilepton asymmetry of 
neutral-B decays. They also discuss  some general effects caused by 
CPT violation. 
 
Neutral-meson interferometry is a powerful tool for investigating   
discrete symmetry. The CP violation ($\epsilon$), direct T and direct 
CP violation are all observed in $K^0 \bar{K^0}$ complex. In the decay 
chain $B\to J/\psi+K \to J/\psi+[f]$, neutral K mixing follows on 
the heels of B mixing. This mixing is called "cascade mixing" and 
the decay is "cascade decay". Cascade mixing has been got some 
theorists' interests \cite{Azimov} \cite{Kayser1}. 
The extension to the decay chain $B\to D \to [f]_D$ for   
exploring new physics can be found in \cite{Meca}. 
The advantage of cascade decay is that we can use the 
known K mixing parameter to  determine the B mixing parameter. 
In \cite{Kayser1}, Kayser shows  
that the cascade decay contains more information than the usually 
discussed processes. So, cascade decay provides a complex and elegant 
window for the detail of the $B^0-\bar{B^0}$ mixing.

In this work, we shall make a detailed study of the CPT violating 
effects caused by $B^0-\bar{B^0}$ mixing in cascade decays. The 
$B^0-\bar{B^0}$ mixing is described by two complex phases $\theta$,  
$\phi$. We first take an approximation method to treat the 
parameters $\theta$ and $\phi$ and give the general formulas of the 
direct CPT and T asymmetry. Then we investigate how to extract the 
$B^0-\bar{B^0}$ mixing parameters. The feasibility of exploring 
the CPT violation in B-factories and LHC-B is discussed. 
Finally, we suggest a procedure to determine the two complex 
phases.

\section*{II.  $B^0-\bar{B^0}$ mixing and CP, T, CPT Asymmetries}

Weak interaction can cause oscillation between $B^0$ and 
$\bar{B^0}$. The eigenstates  of weak decays are not $B^0$ and 
$\bar{B^0}$ but their superpositions which have the simple exponential 
laws. The two eigenstates  of $B^0 \bar B^0$ mesons are given by
\begin{eqnarray}
|B_L>=\frac{1}{\sqrt{|p_1^2|+|q_1|^2}}[p_1|B^0>+q_1|\bar{B^0}>] 
\nonumber\\
|B_H>=\frac{1}{\sqrt{|p_2^2|+|q_2|^2}}[p_2|B^0>-q_2|\bar{B^0}>]
\end{eqnarray}
and their eigenvalues are 
\begin{eqnarray}
\mu_L=m_L-\frac{i}{2}\Gamma_L
     =m_B-\frac{\Delta m_B}{2}-\frac{i}{2}(\Gamma_B+\frac{\Delta \Gamma_B}{2})
     =m_B-\frac{i}{2}\Gamma_B-\frac{\Delta m_B}{2}-\frac{i}{2}y\Gamma_B
\nonumber\\
\mu_H=m_H-\frac{i}{2}\Gamma_H
     =m_B+\frac{\Delta m_B}{2}-\frac{i}{2}(\Gamma_B-\frac{\Delta \Gamma_B}{2})
     =m_B-\frac{i}{2}\Gamma_B+\frac{\Delta m_B}{2}+\frac{i}{2}y\Gamma_B
\end{eqnarray}
From PDG98\cite{PDG98}, $x\equiv\frac{\Delta m_B}{\Gamma_B}\sim 0.7$, 
while $y\equiv\frac{\Delta \Gamma_B}{2\Gamma_B}\leq 10^{-2}$ is 
theoretically expected \cite{Nir1}. 

The mixing parameter $p_i$, $q_i$ are related by\cite{Lee} 
\begin{eqnarray}
\frac{q_1}{p_1}= tg\frac{\theta}{2}e^{i\phi}, ~~~~
\frac{q_2}{p_2}=ctg\frac{\theta}{2}e^{i\phi}  
\end{eqnarray}
where $\theta$ and $\phi$ are complex phases in general.  For real 
$\theta$ and $\phi$,  $0<\theta<\pi$, $0\leq\phi<2\pi$. 

From Eq.(1), The mass difference between $B^0$ and $\bar{B^0}$ is:
\begin{equation}
M_{B^0}-M_{\bar{B^0}}=(\mu_L-\mu_H)\frac{p_1q_2-p_2q_1}{p_1q_2+p_2q_1}
  =(\mu_L-\mu_H)cos\theta
\end{equation}
where $M_{\stackrel{(-)}{B^0}}=m_{\stackrel{(-)}{B^0}}
-\frac{i}{2}\Gamma_{\stackrel{(-)}{B^0}}$.

Using the Bell-Steinberg unitarity relation \cite{Bell}, 
\begin{equation}
|<B_H|B_L>|\leq \frac{\Gamma_L \Gamma_H}{|\mu_L^*-\mu_H|}
  =\frac{\sqrt{1-y^2}\Gamma_B}{|i-x|\Gamma_B}\simeq 0.8
\end{equation}
This constraint is more relaxed than that in $K^0-\bar{K^0}$ complex 
where $|<K_L|K_S>|\leq 0.06$. So,  $|B_H>$ and $|B_L>$ are likely 
unorthogonal. The more relaxed constraint of Eq.(5) perhaps indicates 
the large CP or CPT violation in $B^0$ system. 

The initially $|B^0>$ or $|\bar{B^0}>$ will  evolve after a proper 
time $t$ to  
\begin{eqnarray}
|B^0(t)>=g_+(t)|B^0>+\bar g_+(t)|\bar{B^0}>
\nonumber\\
|\bar B^0(t)>=g_-(t)|\bar{B^0}>+\bar g_-(t)|B^0>
\end{eqnarray}
where 
\begin{eqnarray}
g_+(t)=f_+(t)+cos\theta f_-(t), ~~~&g_-(t)=f_+(t)-cos\theta f_-(t)
\nonumber\\
\bar{g}_+(t)=sin\theta e^{i\phi}f_-(t), ~~~~&
\bar{g}_-(t)=sin\theta e^{-i\phi}f_-(t)
\end{eqnarray}
and
\begin{eqnarray}
f_+(t)=\frac{1}{2}(e^{-i\mu_L t}+e^{-i\mu_H t})
  =e^{-im_B t-\frac{1}{2}\Gamma_B t}ch(\frac{ix-y}{2}\Gamma_Bt)
\nonumber\\
f_-(t)=\frac{1}{2}(e^{-i\mu_L t}-e^{-i\mu_H t})
  =e^{-im_B t-\frac{1}{2}\Gamma_B t}sh(\frac{ix-y}{2}\Gamma_Bt)
\end{eqnarray}

The probability for $|B^0>$ in a proper time $t$ to transform into 
$|B^0>$ is:
\begin{equation}
P_{B^0(t)\to B^0}=|<B^0|B^0(t)>|^2=|g_+(t)|^2
\end{equation}
Similarly, we can define  $P_{B^0(t)\to \bar B^0}$, 
$P_{\bar B^0(t)\to B^0}$, and $P_{\bar B^0(t)\to \bar B^0}$. 

So, the mixing-induced CPT and T asymmetries can be defined as 
\begin{equation}
A_{CPT}(t)\equiv \frac{P_{B^0(t)\to B^0}-P_{\bar B^0(t)\to \bar B^0}}
   {P_{B^0(t)\to B^0}+P_{\bar B^0(t)\to \bar B^0}} 
  =\frac{2Re[cos\theta sh(\frac{ix-y}{2}\Gamma_Bt)
   (ch(\frac{ix-y}{2}\Gamma_Bt))^*]}
   {|ch(\frac{ix-y}{2}\Gamma_Bt)|^2+
    |cos\theta|^2 |sh(\frac{ix-y}{2}\Gamma_Bt)|^2},
\end{equation}
and
\begin{equation}
A_T(t)\equiv \frac{P_{B^0(t)\to \bar B^0}-P_{\bar B^0(t)\to B^0}}
  {P_{B^0(t)\to \bar B^0}+P_{\bar B^0(t)\to B^0}}
  =\frac{|e^{i\phi}|^2-|e^{-i\phi}|^2}
   {|e^{i\phi}|^2+|e^{-i\phi}|^2}
\end{equation}

Some analysis can lead to the following conclusions 
\cite{Sanda2}: \\
(1) CPT invariance requires $cos\theta=0$, and thus 
$\theta=\frac{\pi}{2}$. \\
(2) T invariance requires $\phi=0$; \\
(3) CP conservation requires $cos\theta=0$ ( and thus 
 $\theta=\frac{\pi}{2}$ ) and $\phi=0$. 

In Standard Model, CPT is conserved and
$|\frac{q}{p}|-1=|e^{i\phi}|-1=\frac{1}{2}
Im\frac{\Gamma_{12}}{M_{12}}\sim {\cal O}(10^{-3})$ \cite{Nir2}. 
Thus, the direct T violating 
asymmetry is about the order of ${\cal O}(10^{-3})$. From the 
experience in K system, we guess that the CPT violating effects 
may be smaller than the CP violating effects. 

Under the above assumption, it is convenient to introduce a 
complex $\theta'$ and two real $\phi_0$, $\phi'$ by 
\begin{eqnarray}
\theta=\frac{\pi}{2}+\theta', ~~~&\theta'=Re\theta'+iIm\theta'
\nonumber\\
\phi=\phi_0+i\phi', ~~~&\phi_0=Re\phi, ~\phi'=Im\phi
\end{eqnarray}
where $Re\theta'$, $Im\theta'$, $\phi_0$, $\phi'$ are all real, and 
$|\theta'|<<1$,  ~$|\phi'|<<1$. The relation between $\phi_0$ and 
CKM phase $\beta$ is $\phi_0=-2\beta$.   

Then, we have a very simple relation
\begin{eqnarray}
cos\theta=-\theta', ~~~~sin\theta=1
\nonumber\\
e^{i\phi}=e^{i\phi_0}(1-\phi')~~~~~~~~
\end{eqnarray}
Here we only keep terms up to the first order of $\theta'$ and 
$\phi'$. 

From PDG98 \cite{PDG98}, The mixing parameter
 $\frac{\Delta \Gamma_B}
{\Gamma_B}$ has not been measured experimentally up to now. We further 
assume $y=0$ in order to simplify the formulations below. Thus, 
\begin{equation}
ch(\frac{ix-y}{2}\Gamma_Bt)=cos\frac{\Delta m_B t}{2}, ~~~~~~~~~ 
sh(\frac{ix-y}{2}\Gamma_Bt)=isin\frac{\Delta m_B t}{2}
\end{equation}

From Eq.(10), Eq.(11),  Eq.(13) and Eq.(14), we obtain 
\begin{eqnarray}
A_{CPT}(t)&=&\frac{2Im\theta' sin\Delta m_B t}{1+cos\Delta m_B t}
\nonumber\\
A_{T}(t)&=&-2\phi' 
\end{eqnarray}
So, the mixing-induced CPT asymmetry is proportional to $Im\theta'$, and 
the mixing-induced T asymmetry is proportional to $Im\phi=\phi'$.

Now, we discuss two cases:

(1) Final state is not CP eigenstate

We study the decays $B^0\to Xl^+\nu$, $\bar{B^0}\to\bar Xl^-\nu$.
From $\Delta B=\Delta Q$ rule, the decays of $B^0\to\bar Xl^-\nu$,
$\bar{B^0}\to Xl^+\nu$ are forbidden. 

For the allowed processes, we define the amplitude:
\begin{eqnarray}
<Xl^+\nu|H|B^0>=A, ~~~~~~~<\bar X l^-\nu|H|\bar{B^0}>=A^*
\nonumber
\end{eqnarray}
So the asymmetry of semileptonic decay rates are 
\begin{eqnarray}
D_1(f,t)&\equiv& \frac
  {\Gamma(B^0(t)\to Xl^+\nu)-\Gamma(\bar{B^0}(t)\to \bar X l^-\nu)}
  {\Gamma(B^0(t)\to Xl^+\nu)+\Gamma(\bar{B^0}(t)\to \bar X l^-\nu)}
  =\frac
  {P_{B^0(t)\to B^0}-P_{\bar{B^0}(t)\to \bar{B^0}}}
  {P_{B^0(t)\to B^0}+P_{\bar{B^0}(t)\to \bar{B^0}}}
  =A_{CPT}(t)
\nonumber\\
D_2(f,t)&\equiv& \frac
  {\Gamma(B^0(t)\to \bar X l^-\nu)-\Gamma(\bar{B^0}(t)\to Xl^+\nu)}
  {\Gamma(B^0(t)\to \bar X l^-\nu)+\Gamma(\bar{B^0}(t)\to Xl^+\nu)}
  =\frac
  {P_{B^0(t)\to \bar{B^0}}-P_{\bar{B^0}(t)\to B^0}}
  {P_{B^0(t)\to \bar{B^0}}+P_{\bar{B^0}(t)\to B^0}}
  =A_{T}(t)
\end{eqnarray}

Eq.(16) shows that CPT and T asymmetry can lead to CP asymmetry. One can 
use the CP asymmetry of semileptonic B decays to measure the CPT and T 
violation parameter.

(2) Final state is CP eigenstate

The decay rate for an initial $B^0$ or $\bar{B^0}$ transform into a
CP eigenstate $f$ is
\begin{eqnarray}
\Gamma(B^0(t)\to f)=e^{-\Gamma_B t}|A|^2
  \{\frac{1+cos\Delta m_Bt}{2}+Im\theta'sin\Delta m_Bt+
  |\frac{\bar A}{A}|^2(1-2\phi')\frac{1-cos\Delta m_Bt}{2}
\nonumber\\
  -Im[e^{i\phi_0}\frac{\bar A}{A}
  (sin\Delta m_Bt-\phi' sin\Delta m_Bt
  +i\theta'^*(1-cos\Delta m_Bt))]\}~~~~
\\
\Gamma(\bar{B^0}(t)\to f)=e^{-\Gamma_B t}|A|^2
  \{(1+2\phi')\frac{1-cos\Delta m_Bt}{2}+|\frac{\bar A}{A}|^2
  (\frac{1+cos\Delta m_Bt}{2}-Im\theta'sin\Delta m_Bt)
\nonumber\\
  +Im[e^{i\phi_0}\frac{\bar A}{A}
  (sin\Delta m_Bt+\phi' sin\Delta m_Bt
  +i\theta'(1-cos\Delta m_Bt))]\}~~~~
\nonumber
\end{eqnarray}
where $A\equiv A(B^0\to f)$ and $\bar A\equiv A(\bar{B^0}\to f)$.  

For $f=J/\psi K_S$, $\frac{\bar A}{A}=-1$, the CP asymmetry is
\begin{eqnarray}
D(J/\psi K_S,t)&=&\frac
 {\Gamma(B^0(t)\to J/\psi K_S)-\Gamma(\bar{B^0}(t)\to J/\psi K_S)}
 {\Gamma(B^0(t)\to J/\psi K_S)+\Gamma(\bar{B^0}(t)\to J/\psi K_S)}
\nonumber\\
&=&sin\phi_0sin\Delta m_Bt+Re\theta'cos\phi_0(1-cos\Delta m_Bt)
 +Im\theta'sin\Delta m_Bt
\nonumber\\ 
 &&-Im\theta'sin\phi_0(1-cos\Delta m_Bt)
 -\phi'(1-cos\Delta m_Bt)+\phi'sin\phi_0sin\Delta m_Bt
\end{eqnarray}

For $f=J/\psi K_L$, $\frac{\bar A}{A}=+1$, the CP asymmetry is
\begin{eqnarray}
D(J/\psi K_L,t)&=&\frac
 {\Gamma(B^0(t)\to J/\psi K_L)-\Gamma(\bar{B^0}(t)\to J/\psi K_L)}
 {\Gamma(B^0(t)\to J/\psi K_L)+\Gamma(\bar{B^0}(t)\to J/\psi K_L)}
\nonumber\\
&=&-sin\phi_0sin\Delta m_Bt-Re\theta'cos\phi_0(1-cos\Delta m_Bt)
 +Im\theta'sin\Delta m_Bt
\nonumber\\ 
 &&+Im\theta'sin\phi_0(1-cos\Delta m_Bt)
 -\phi'(1-cos\Delta m_Bt)-\phi'sin\phi_0sin\Delta m_Bt
\end{eqnarray}
    
\section*{III.  Cascade decays}
We have discussed the $B^0-\bar{B^0}$ mixing in the previous section. 
Now we turn to the cascade mixing involving both neutral B and neutral 
K systems in succession. Neglecting CPT violating effects    
in the neutral K system, the weak eigenstates of the 
neutral K mesons can be represented by the usual form:
\begin{eqnarray}
|K_S>=\frac{1}{\sqrt{|p_K^2|+|q_K|^2}}[p_K|B^0>+q_K|\bar{B^0}>]
\nonumber\\
|K_L>=\frac{1}{\sqrt{|p_K^2|+|q_K|^2}}[p_K|B^0>-q_K|\bar{B^0}>]
\end{eqnarray}
and their eigenvalues are
\begin{eqnarray}
\mu_{S(L)}=m_K{\stackrel{(+)}{-}}\frac{\Delta m_K}{2}
           -i\frac{\Gamma_{S(L)}}{2}
\end{eqnarray}
where $m_K$ is the average of the $K_S$ and $K_L$ masses, 
$\Gamma_{S,L}$ are the $K_{S,L}$ widths. 

The time evolution of the neutral K mesons can be easily obtained 
\begin{eqnarray}
|K^0(t)>=e_+(t)|K^0>+\frac{q_K}{p_K}e_-(t)|\bar{K^0}>
\nonumber\\ 
|\bar{K^0}(t)>=\frac{p_K}{q_K}e_-(t)|K^0>+e_+(t)|\bar{K^0}> 
\end{eqnarray}
where 
\begin{eqnarray}
e_{\pm}(t)=\frac{1}{2}(e^{-i\mu_S}\pm e^{-i\mu_L})
\end{eqnarray}

Consider the decay chain $B\to J/\psi+K \to J/\psi+[f]$  
where $f$ can be $2\pi$, $3\pi$ and $\pi l\nu$ as shown in 
Fig.1. Other decay 
modes of the neutral K mesons are neglected because of either 
very small fractions or less physical interest 
for this paper. We first give the formulations of the 
most complicated case where the final state $f=\pi l\nu$.

According to \cite{Kayser2}, the decay amplitude of 
the cascade decay 
$B^0{\stackrel{t_1}{\to}} J/\psi+K{\stackrel{t_2}{\to}} 
J/\psi+[\pi^-l^+\nu]$ is
\begin{eqnarray}
&& A(B^0{\stackrel{t_1}{\to}} J/\psi+K{\stackrel{t_2}{\to}} 
    J/\psi+[\pi^-l^+\nu])=
    g_+(t_1)A(B^0\to J/\psi K^0)e_+(t_2)A(K^0\to \pi^-l^+\nu)
\nonumber\\
&& ~~~~~~~ +\bar g_+(t_1)A(\bar{B^0}\to J/\psi \bar{K^0})
   \frac{p_K}{q_K}e_-(t_2)A(K^0\to \pi^-l^+\nu) 
\end{eqnarray}

We assume that the transition amplitude for B and K decays satisfy 
the $\Delta S=\Delta Q$ rule and CP, CPT invariance. There is no 
experimental signal of the violation $\Delta S=\Delta Q$ rule. 
The assumption of CP conservation in the transition amplitude for 
B and K decays is valid because 
$\frac{A(\bar{B^0}\to J/\psi \bar{K^0})}{A(B^0\to J/\psi K^0)}=-1$ 
to a very high degree and the direct CP violation in 
$K^0-\bar{K^0}$ system is very small 
($Re(\epsilon'/\epsilon)\sim 10^{-3}$). 
We further neglect the small CP violations in $ K^0-\bar{K^0}$ 
mixing, thus $\frac{q_K}{p_K}=1$. 

Under the above assumptions, the decay rate of the cascade decay 
$B^0{\stackrel{t_1}{\to}} J/\psi+K{\stackrel{t_2}{\to}} 
J/\psi+[\pi^{\mp}l^{\pm}\nu]$ is 
\begin{eqnarray} 
&&\Gamma(B^0, J/\psi+[\pi^{\mp}l^{\pm}\nu])\equiv
  \Gamma(B^0{\stackrel{t_1}{\to}} J/\psi+K{\stackrel{t_2}{\to}} 
   J/\psi+[\pi^{\mp}l^{\pm}\nu])
\nonumber\\
&&\propto e^{-\Gamma_Bt_1}\{e^{-\Gamma_St_2}
  [1+sin\phi_0sin\Delta m_Bt_1+Re\theta'cos\phi_0(1-cos\Delta m_Bt_1)
   +Im\theta'sin\Delta m_Bt_1
\nonumber\\
&& +Im\theta'sin\phi_0(1-cos\Delta m_Bt_1)
   -\phi'(1-cos\Delta m_Bt_1)-\phi'sin\phi_0sin\Delta m_Bt_1] 
\nonumber\\
&&+e^{-\Gamma_Lt_2}
  [1-sin\phi_0sin\Delta m_Bt_1-Re\theta'cos\phi_0(1-cos\Delta m_Bt_1)
   +Im\theta'sin\Delta m_Bt_1
\nonumber\\
&& -Im\theta'sin\phi_0(1-cos\Delta m_Bt_1)
   -\phi'(1-cos\Delta m_Bt_1)+\phi'sin\phi_0sin\Delta m_Bt_1] 
\nonumber\\
&& \pm 2e^{-\frac{1}{2}(\Gamma_S+\Gamma_L)t_2}
  [cos\Delta m_Bt_1cos\Delta m_Kt_2+cos\phi_0sin\Delta m_Bt_1sin\Delta m_Kt_2
\nonumber\\ 
&& -Re\theta'sin\phi_0(1-cos\Delta m_Bt_1)
   +Im\theta'sin\Delta m_Bt_1cos\Delta m_Kt_2
\nonumber\\
&& +Im\theta'cos\phi_0(1-cos\Delta m_Bt_1)sin\Delta m_Kt_2
   +\phi'(1-cos\Delta m_Bt_1)cos\Delta m_Kt_2
\nonumber\\
&& -\phi'cos\phi_0sin\Delta m_Bt_1sin\Delta m_Kt_2]
  \}
\end{eqnarray}

Similarly, the decay rate of the cascade decay 
$\bar{B^0}{\stackrel{t_1}{\to}} J/\psi+K{\stackrel{t_2}{\to}} 
J/\psi+[\pi^{\pm}l^{\mp}\nu]$ is 
\begin{eqnarray}
&&\Gamma(\bar B^0, J/\psi+[\pi^{\pm}l^{\mp}\nu])\equiv
  \Gamma(\bar{B^0}{\stackrel{t_1}{\to}} J/\psi+K{\stackrel{t_2}{\to}} 
    J/\psi+[\pi^{\pm}l^{\mp}\nu])
\nonumber\\
&& =\Gamma(B^0{\stackrel{t_1}{\to}} J/\psi+K{\stackrel{t_2}{\to}} 
  J/\psi+[\pi^{\mp}l^{\pm}\nu])
  (\theta'\to -\theta', \phi_0\to -\phi_0, \phi'\to -\phi')
\end{eqnarray}

Because we have neglected the small CP-violating effects in K system, 
only $K_S\to 2\pi$ and  $K_L\to 3\pi$ are possible. The decay rate 
for the cascade decays of the $f=2\pi$ and $f=3\pi$ are:
\begin{eqnarray}
&&\Gamma({\stackrel{(-)}{B^0}}, J/\psi+[2\pi])\equiv
  \Gamma({\stackrel{(-)}{B^0}}{\stackrel{t_1}{\to}} J/\psi+K_S
   {\stackrel{t_2}{\to}}J/\psi+[2\pi])
\nonumber\\
&&\propto 4e^{-\Gamma_Bt_1}\{e^{-\Gamma_St_2}
  [1{\stackrel{(-)}{+}}sin\phi_0sin\Delta m_Bt_1
   {\stackrel{(-)}{+}}Re\theta'cos\phi_0(1-cos\Delta m_Bt_1)
   {\stackrel{(-)}{+}}Im\theta'sin\Delta m_Bt_1
\nonumber\\
&&~~~~~ +Im\theta'sin\phi_0(1-cos\Delta m_Bt_1)
   {\stackrel{(+)}{-}}\phi'(1-cos\Delta m_Bt_1)
   -\phi'sin\phi_0sin\Delta m_Bt_1]
 \}  
\end{eqnarray}
and 
\begin{eqnarray}
&&\Gamma({\stackrel{(-)}{B^0}}, J/\psi+[3\pi])\equiv
  \Gamma({\stackrel{(-)}{B^0}}{\stackrel{t_1}{\to}} J/\psi+K_L
   {\stackrel{t_2}{\to}}J/\psi+[3\pi])
\nonumber\\
&&\propto 4e^{-\Gamma_Bt_1}\{e^{-\Gamma_St_2}
  [1{\stackrel{(+)}{-}}sin\phi_0sin\Delta m_Bt_1
   {\stackrel{(+)}{-}}Re\theta'cos\phi_0(1-cos\Delta m_Bt_1)
   {\stackrel{(-)}{+}}Im\theta'sin\Delta m_Bt_1
\nonumber\\
&&~~~~~ +Im\theta'sin\phi_0(1-cos\Delta m_Bt_1)
   {\stackrel{(+)}{-}}\phi'(1-cos\Delta m_Bt_1)
   +\phi'sin\phi_0sin\Delta m_Bt_1]
 \}  
\end{eqnarray}

\section*{IV.  The determination of the parameter
$\theta$ and $\phi$}
1. $\phi'$

From Eq.(16) and \cite{Sanda1}, 
\begin{eqnarray}
A_{T}(t)=\frac{\Gamma(B^0(t)\to \bar Xl^-\nu)-
   \Gamma(\bar{B^0}(t)\to Xl^+\nu)}
  {\Gamma(B^0(t)\to \bar Xl^-\nu)+
   \Gamma(\bar{B^0}(t)\to Xl^+\nu)}
  =\frac{N^{++}-N^{--}}{N^{++}+N^{--}}
  =-2\phi'
\end{eqnarray}
where $N^{++}$, $N^{--}$ are the same-sign dilepton events. 
  
$\phi'$ can be measured by the semileptonic decays of the B mesons 
or by the same-sign dilepton ratios suggested in \cite{Sanda1}.

\noindent
2. $Im\theta'$

From Eq.(16) and \cite{Sanda1}, 
\begin{eqnarray}
A_{CPT}(t)=\frac{\Gamma(B^0(t)\to Xl^+\nu)-
   \Gamma(\bar{B^0}(t)\to \bar Xl^-\nu)}
  {\Gamma(B^0(t)\to Xl^+\nu)+
   \Gamma(\bar{B^0}(t)\to \bar Xl^-\nu)}
  =\frac{N^{+-}-N^{-+}}{N^{+-}+N^{-+}}
  =\frac{2Im\theta'sin\Delta m_Bt}{1+cos\Delta m_Bt}
\end{eqnarray}
where $N^{+-}$, $N^{-+}$ are opposite-sign dilepton events. 

Thus, $Im\theta'$ can be measured by the semileptonic decays of
the B mesons or by the opposite-sign dilepton ratios.  For the dileptonic
decays in Eq.(29) and Eq.(30), they correspond to the case of $C=-1$
where $C$ is the charge conjugation number of the $B^0\bar{B^0}$ pair. 

Another method is: from Eq.(27) and Eq.(28),
\begin{eqnarray}
&&\frac{\Gamma(B^0, J/\psi+[2\pi])-\Gamma(\bar{B^0}, J/\psi+[2\pi])}
   {\Gamma(B^0, J/\psi+[2\pi])+\Gamma(\bar{B^0}, J/\psi+[2\pi])}
 +\frac{\Gamma(B^0, J/\psi+[3\pi])-\Gamma(\bar{B^0}, J/\psi+[3\pi])}
    {\Gamma(B^0, J/\psi+[3\pi])+\Gamma(\bar{B^0}, J/\psi+[3\pi])}
\nonumber\\
&&=2[Im\theta'sin\Delta m_Bt_1-\phi'(1-cos\Delta m_Bt_1)]
 \end{eqnarray}
Using the known $\phi'$ value from the semileptonic decays or the
dilepton ratios, $Im\theta'$ can be determined by Eq.(31).  
But this method is not good for experiment because  their branching 
ratios are smaller than the  semileptonic decays.

\noindent
3. $sin\phi_0$ and the absolute value of the 
$cos\phi_0$ and $Re\theta'$

From Eq.(27) and (28), we can obtain the time-dependent asymmetry of 
the decay rates 
\begin{eqnarray}
&&\frac
  {[\Gamma(B^0, J/\psi+[2\pi])-\Gamma(\bar{B^0}, J/\psi+[2\pi])]}
  {[\Gamma(B^0, J/\psi+[2\pi])+\Gamma(\bar{B^0}, J/\psi+[2\pi])]}
 -\frac
  {[\Gamma(B^0, J/\psi+[3\pi])-\Gamma(\bar{B^0},J/\psi+[3\pi])]}
  {[\Gamma(B^0, J/\psi+[3\pi])+\Gamma(\bar{B^0},J/\psi+[3\pi])]}
  \nonumber\\
&&=2[sin\phi_0sin\Delta m_Bt_1+
   Re\theta'cos\phi_0(1-cos\Delta m_Bt_1)]
   \end{eqnarray}

The asymmetry of Eq.(32) is twice as much as the
usual CP asymmetry in the decay of $B\to J/\psi K_S$ because we
have used the decay mode of $B\to J/\psi K_L$ to double the asymmetry.
We will discuss the problem caused by the detection of $K_L$ later.

There are two contributions in the asymmetry of Eq.(32). The
$sin\Delta m_Bt_1$ term is an odd function of time while the
$(1-cos\Delta m_Bt_1)$ term is an even function. These two terms can be
distinguished experimentally by measuring the decay time order of
$B^0$ abd $\bar{B^0}$
decays. The detail of this method is given in \cite{Gronau}. Here we
only use this method to distinguish the $sin\Delta m_Bt_1$ and
$(1-cos\Delta m_Bt_1)$ terms. 

Like \cite{Gronau}, define two asymmetries 
\begin{eqnarray*}
a_{-}(f,t)=\frac
    {(\bar \Gamma+{\stackrel{\sim }{\Gamma}})
    -(\Gamma+{\stackrel{\sim }{\bar \Gamma}})}
    {(\bar \Gamma+{\stackrel{\sim }{\Gamma}}) 
    +(\Gamma+{\stackrel{\sim }{\bar \Gamma}})} ,~~~~
a_{+}(f,t)=\frac
    {(\bar \Gamma+{\stackrel{\sim }{\bar \Gamma}})
    -(\Gamma+\bar \Gamma)}
    {(\bar \Gamma+{\stackrel{\sim }{\bar \Gamma}})
    +(\Gamma+\bar\Gamma)}
\end{eqnarray*}
where $\Gamma$, $\bar \Gamma$, ${\stackrel{\sim }{\Gamma}}$ and 
${\stackrel{\sim }{\bar \Gamma}}$ are defined in \cite{Gronau},  
and the subscript (-) or (+) denotes an odd or even function of time.
The measurement of asymmetry $a_{-}(f,t)$ requires measuring the decay 
time order. 

Thus the two terms of Eq.(32) can be distinguished by
\begin{eqnarray}
A_{-}(t_1)=a_{-}(f_1,t_1)-a_{-}(f_2,t_1)=2sin\phi_0sin\Delta m_Bt_1
\end{eqnarray}
and 
\begin{eqnarray}
A_{+}(t_1)=a_{+}(f_1,t_1)-a_{+}(f_2,t_1)
 =2Re\theta'cos\phi_0(1-cos\Delta m_Bt_1)
\end{eqnarray}
where $f_1$ and $f_2$ represent the final states $J/\psi+[2\pi]$ and
$J/\psi+[3\pi]$. 

In Eq.(34), the asymmetry is often used to measure the direct CP violation 
when CPT invariance holds. The direct CP violation in $B\to J/\psi K$ 
decays is at the order of  ${\cal O}(10^{-3})$. So the measurement of 
the CPT violation in Eq.(34) can only reach the accuracy up to  
$10^{-2}$ because of the dilution of direct CP violation 
in $B\to J/\psi K$ decay and the CP violation in $K^0-\bar{K^0}$ mixing. 
 
From the Eq.(33) and Eq.(34), the values of $sin\phi_0$
and $Re\theta'cos\phi_0$ can be obtained. So 
$cos\phi_0$ and $Re\theta'$ can be determined 
except for the ambiguity of their sign. This ambiguity can be
solved by the cascade decays where $f=\pi l\nu$. 

\noindent
4. The sign of $cos\phi_0$ and $Re\theta'$

From Eq.(25) and Eq.(26), 
\begin{eqnarray}
&&\frac
  {[\Gamma(     B^0,  J/\psi+[\pi^- l^+\nu])
   -\Gamma(     B^0,  J/\psi+[\pi^+ l^-\nu])
   +\Gamma(\bar{B^0}, J/\psi+[\pi^- l^+\nu])
   -\Gamma(\bar{B^0}, J/\psi+[\pi^+ l^-\nu])}
  {[\Gamma(     B^0,  J/\psi+[\pi^- l^+\nu])
   +\Gamma(     B^0,  J/\psi+[\pi^+ l^-\nu])
   +\Gamma(\bar{B^0}, J/\psi+[\pi^- l^+\nu])
   +\Gamma(\bar{B^0}, J/\psi+[\pi^+ l^-\nu])}
\nonumber\\
&&=\frac{-2e^{-\frac{1}{2}(\Gamma_S+\Gamma_L)t_2}
         Re\theta'sin\phi_0(1-cos\Delta m_Bt_1)}
        {\{e^{-\Gamma_St_2}[1+Im\theta'sin\phi_0(1-cos\Delta m_Bt_1)
	 -\phi'sin\phi_0sin\Delta m_Bt_1]}
\nonumber\\
&&~~    +e^{-\Gamma_Lt_2}[1-Im\theta'sin\phi_0(1-cos\Delta m_Bt_1)
	 +\phi'sin\phi_0sin\Delta m_Bt_1]\}
\\
&& \sim \frac{-2e^{-\frac{1}{2}(\Gamma_S+\Gamma_L)t_2}
         Re\theta'sin\phi_0(1-cos\Delta m_Bt_1)}
	{e^{-\Gamma_St_2}+e^{-\Gamma_Lt_2}} \nonumber
\end{eqnarray}
and 
\begin{eqnarray}
&&\frac
  {[\Gamma(     B^0,  J/\psi+[\pi^- l^+\nu])
   -\Gamma(     B^0,  J/\psi+[\pi^+ l^-\nu])
   -\Gamma(\bar{B^0}, J/\psi+[\pi^- l^+\nu])
   +\Gamma(\bar{B^0}, J/\psi+[\pi^+ l^-\nu])}
  {[\Gamma(     B^0,  J/\psi+[\pi^- l^+\nu])
   +\Gamma(     B^0,  J/\psi+[\pi^+ l^-\nu])
   +\Gamma(\bar{B^0}, J/\psi+[\pi^- l^+\nu])
   +\Gamma(\bar{B^0}, J/\psi+[\pi^+ l^-\nu])}
\nonumber\\
&&=\frac {A}{B} \sim \frac
   { 2e^{-\frac{1}{2}(\Gamma_S+\Gamma_L)t_2}
         [cos\Delta m_Bt_1cos\Delta m_Kt_2
	 +cos\phi_0sin\Delta m_Kt_2sin\Delta m_Bt_1]}
   {e^{-\Gamma_St_2}+e^{-\Gamma_Lt_2}}	  
\end{eqnarray}
where
\begin{eqnarray}
A&=&    2e^{-\frac{1}{2}(\Gamma_S+\Gamma_L)t_2}
	 [cos\Delta m_Bt_1cos\Delta m_Kt_2 
	 +cos\phi_0sin\Delta m_Kt_2(sin\Delta m_Bt_1 \nonumber\\
   &&      -Im\theta'(1-cos\Delta m_Bt_1)    
        +\phi'sin\Delta m_Bt_1)  \nonumber\\ 
   &&      -Im\theta'sin\Delta m_Bt_1cos\Delta m_Kt_2
         -\phi'(1-cos\Delta m_Bt_1)]  \nonumber\\
B&=&    e^{-\Gamma_St_2}[1+Im\theta'sin\phi_0(1-cos\Delta m_Bt_1)
         -\phi'sin\phi_0sin\Delta m_Bt_1] \nonumber\\ 
   &&   +e^{-\Gamma_Lt_2}[1-Im\theta'sin\phi_0(1-cos\Delta m_Bt_1)
     	   +\phi'sin\phi_0sin\Delta m_Bt_1] \nonumber
\end{eqnarray}

From Eq.(35) and Eq.(36), the sign of $cos\phi_0$ and $Re\theta'$ 
can be measured. Actually, we do not need to use the Eq.(35) since 
we have known the value of  $Re\theta'cos\phi_0$.  

\section*{V. Feasibility and discussions}
In order to meet the goal of a three standard deviation measurement 
for the $B^0-\bar{B^0}$ mixing parameter $\theta$ and $\phi$, the 
relation between the number of $B^0\bar{B^0}$ pairs and the 
asymmetry is \cite{SLAC89}: 
\begin{eqnarray}
N_{B^0\bar{B^0}}=\frac{1}{B\epsilon_r\epsilon_t
 [(1-2W)d\cdot\delta A]^2}
\end{eqnarray}
where $\delta A=\frac{A}{3}$; $A$ is the asymmetry of the decay 
ratios; $B$ is the branching ratio of the  decay; 
$\epsilon_r$ is the reconstruction efficiency of the final 
state $f$; $\epsilon_t$ is the tagging efficiency; 
$W$ is the fraction of incorrect tags; 
$d$ is the dilution factor which takes into account the loss 
in asymmetry due to fitting, time integration, and/or the mixing 
of the tagged decay.  

Sometimes, another relation is used: 
\begin{eqnarray}
N_{eff}=\frac{1}{[(1-2W)d\cdot\delta A]^2}
\end{eqnarray}
where $N_{eff}$ is the effective number of the decay events.  

The  B-factories can accumulate  $1.8\times 10^8$ $B\bar B$ 
pairs every year \cite{PEP}, and the effective event number  of 
Eq.(38) for $B\to J/\psi K_S$ is taken to be 
$2.7\times 10^{5}$ in LHC-B\cite{LHC}. Table 1 and Table 2 give some  
parameters \cite{SLAC89} and the minimum asymmetries (lower bound) 
which can be achieved in  B-factory and LHC-B.

\begin{table}[hbt]
\begin{center} 

Table 1. Branching Ratios and Reconstruction efficiencies \\ 
  for the cascade decays and semileptonic decays. 

\vspace{0.6cm}
\begin{tabular}{|c|c|c|} \hline \hline
Decay mode & Braching Ratio $B$ & Reconstruction effeciency  
$\epsilon_r$ \\ \hline
 
$B\to J/\psi K_S \to J/\psi+[2\pi]$  & $5\times 10^{-4}$ 
&0.61  \\ \hline

$B\to J/\psi K_L \to J/\psi+[3\pi]$  
& $5\times 10^{-4}\times \frac{1}{3}$ & 0.4 \\ \hline

$B\to J/\psi K_S \to J/\psi+[\pi l\nu]$  
& $5\times 10^{-4} \cdot 1.2\times 10^{-3}$ 
& 0.61\\ \hline

$B\to J/\psi K_L \to J/\psi+[\pi l\nu]$  
& $5\times 10^{-4} \times \frac{2}{3}$ & 0.4 \\ \hline

$B\to l\nu+X$ & 0.1 & 1 \\ \hline
$J/\psi\to l^+l^-$ & 0.14 & - \\ \hline

\end{tabular}

\vspace{0.8cm}
Tag efficiencies and Asymmetry dilution at B-factory and LHC-B 

\vspace{0.5cm}
\begin{tabular}{|c|c|c|} \hline \hline
& at B-factory  & at LHC-B \\ \hline 
Tag efficiency $\epsilon_t$ & 0.48 & 0.61 \\ \hline
Asymmetry dilution $d$ & 0.61(for $B_d$) &0.61 \\ \hline

\end{tabular}
\end{center}
\end{table}

\begin{table}[hbt]
\begin{center}

Table 2. Comparison between the Minimum of Asymmetries \\
 with $3\sigma$ standard deviation at B-factory and LHC-B  

\vspace{0.5cm}
\begin{tabular}{|c|c|c|} \hline \hline
Decay mode & \multicolumn{2}{|c|} {Asymmetry $A$} \\\cline{2-3} 
& at B-factory  & at LHC-B \\ \hline

$B\to J/\psi K_S \to J/\psi+[2\pi]$  
& 0.08 & $1.2\times 10^{-2}$   \\ \hline

$B\to J/\psi K_L \to J/\psi+[3\pi]$  
& 0.17 & $2.6\times 10^{-2}$   \\ \hline

$B\to J/\psi K_S \to J/\psi+[\pi l\nu]$  
&  1.44 & 0.27\\ \hline

$B\to J/\psi K_L \to J/\psi+[\pi l\nu]$  
& 0.06 & 0.01 \\ \hline

$B\to l\nu+X$ & $1.7\times10^{-3}$ 
& $1\times10^{-5}$ \\ \hline

$B\bar B\to l^+l^-$ & $7\times10^{-3}$ 
& $4\times10^{-5}$ \\ \hline

\end{tabular}
\end{center}
\end{table}

Now we discuss how to determine the two complex phase 
$\theta$ and $\phi$. 

(1) $\phi'$ and $Im\theta'$ can be measured in semileptonic decays 
and dileptonic decays given in Eq.(29) and Eq.(30). Semileptonic 
decays have larger branching ratio but smaller detection efficiency, 
while dileptonic decays have larger detection efficiency but smaller 
branching ratio than semileptonic decays. 
In B-factory (with $1.8\times 10^8$ $B^0\bar{B^0}$ 
pairs per year), the  $\phi'$ and $Im\theta'$ can be measured to 
an accuracy of $2 \times 10^{-3}$ for semileptonic decays 
and $7\times 10^{-3}$ for dileptonic decays at $3\sigma$ level.  
In LHC-B (with $4\times 10^{12}$ $b\bar b$ pairs 
every snow mass year), the  $\phi'$ and $Im\theta'$ can be measured to  
an accuracy of $10^{-5}$ for semileptonic decays and 
$4\times 10^{-5}$ for dileptonic decays with $3\sigma$ standard 
deviation. 

(2) $Re\theta'cos\phi_0$ can be determined clearly from Eq.(34). 
In Eq.(32) the decay of $K_L\to 3\pi$ is used in order to increase 
the asymmetry factor and reduce the ambiguity or error caused 
by the unknown $\phi'$ and $Im\theta'$. The $K_L$ detection is a 
challenge for experiment. In \cite{SLAC92}, one idea to catch $K_L$ 
by using Fe sampling after the electromagnetic calorimeter is 
suggested.  In B-factory,  the accuracy of measuring 
$Re\theta'cos\phi_0$ can reach $0.07$. So, it is likely that 
$Re\theta'cos\phi_0$ can not be measured in B-factory. In LHC-B,  
the accuracy of measuring $Re\theta'cos\phi_0$ can reach $0.01$. 
We have no confidence that $Re\theta'cos\phi_0$ can be measured 
clearly with $3\sigma$ standard deviation in LHC-B. 
If it can be measured, $Re\theta'$ must be larger than $0.01$. 
This will be the largest CPT violation effects. 

(3) $sin\phi_0$ is usually suggested to be measured in 
$B\to J/\psi K_S(\to\pi^+\pi^-)$ which gives the 
$A(t)=sin\phi_0sin\Delta m_Bt$ in Standard Model. If the CPT 
violating effects are considered, the asymmetry is modified to 
Eq.(18). In order to cancel the errors caused by $\phi'$ and 
$Im\theta'$, we use the Eq.(33) to determine $sin\phi_0$. 

(4) $sin\phi_0$  can be measured in B-factory and LHC-B as discussed 
above, but the value of $\phi_0$ has two ambiguity. 
If  $Re\theta'cos\phi_0$ can be measured, the only ambiguity 
is the sign of $cos\phi_0$. This can be solved by measuring the 
ratios of the cascade decays $B\to J/\psi+K\to J/\psi+[\pi l\nu]$ 
given in Eq.(36). Because the very small branching ratio of decay 
$B\to J/\psi+K\to J/\psi+[\pi l\nu]$ in $t_2 \le 2\tau_S$, 
the determination of the sign of $cos\phi_0$ can only be done 
in LHC-B. Table 2 gives the lower bound of measuring for 
asymmetry in cascade decay is 0.2. This is possible because 
$|cos\phi_0|>0.5$, we have taken 
$0.3\leq sin\phi_0\leq 0.88$ \cite{Nir2} for our estimation. 

In conclusion, the cascade decays provide an elegant and 
beautiful place to study the CPT violation caused by the 
$B^0-\bar{B^0}$ mixing.

\section*{Acknowledgment}

This work is supported in part by National Natural Science 
Foundation of China and the Grant of State Commission of Science 
and Technology of China.

\newpage
\begin{figure}
\unitlength=0.5mm

\begin{picture}(100,50)
\put(60,150){\vector(1,1){22}}
\put(60,145){\vector(1,-1){22}}
\put(97,172){\vector(1,0){30}}
\put(97,123){\vector(1,0){30}}
\put(154,172){\vector(1,0){30}}
\put(154,123){\vector(1,0){30}}
\put(155,167){\vector(3,-4){29}}
\put(155,128){\vector(3,4){29}}
\put(210,172){\vector(1,-1){20}}
\put(210,123){\vector(1,1){20}}
\put(49,142){${\stackrel{(-)}{B^0}}$}
\put(64,161){$t_1$}
\put(64,131){$t_1$}
\put(85,170){$B^0$}
\put(85,120){$\bar{B^0}$}
\put(128,170){$J/\psi K^0$}
\put(128,120){$J/\psi \bar{K^0}$}
\put(165,175){$t_2$}
\put(165,115){$t_2$}
\put(186,170){$J/\psi K^0$}
\put(186,120){$J/\psi \bar{K^0}$}
\put(232,145){$J/\psi [f]$}
\end{picture}

\begin{picture}(100,100)
\put(60,150){\vector(1,1){22}}
\put(60,145){\vector(1,-1){22}}
\put(97,172){\vector(1,0){30}}
\put(97,123){\vector(1,0){30}}
\put(154,172){\vector(1,0){30}}
\put(154,123){\vector(1,0){30}}
\put(97,167){\vector(3,-4){29}}
\put(97,128){\vector(3,4){29}}
\put(210,172){\vector(1,-1){20}}
\put(210,123){\vector(1,1){20}}
\put(49,142){${\stackrel{(-)}{B^0}}$}
\put(64,161){$t_1$}
\put(64,131){$t_1$}
\put(85,170){$B_L$}
\put(85,120){$B_H$}
\put(128,170){$J/\psi K_S$}
\put(128,120){$J/\psi K_L$}
\put(165,175){$t_2$}
\put(165,115){$t_2$}
\put(186,170){$J/\psi K_S$}
\put(186,120){$J/\psi K_L$}
\put(232,145){$J/\psi [f]$}
\put(50,80){Fig.1~~ The cascade decay chains  
  $B\to J/\psi K\to J/\psi [f]$ }
\end{picture}

\end{figure}

\end{document}